# Observation of Large Topologically Trivial Fermi-Arcs in the Candidate Type-II Weyl Semimetal WTe$_2$

F. Y. Bruno[1], A. Tamai[1], Q. S. Wu[2], I. Cucchi[1], C. Barreteau[1], A. de la Torre[1], S. McKeown Walker[1], S. Riccò[1], Z. Wang[1,3], T. K. Kim[4], M. Hoesch[4], M. Shi[3], N. C. Plumb[3], E. Giannini[1], A. A. Soluyanov[2,5], and F. Baumberger[1,3,]

[1] *Department of Quantum Matter Physics, University of Geneva, 24 Quai Ernest-Ansermet, 1211 Geneva 4, Switzerland*

[2] *Theoretical Physics and Station Q Zurich, ETH Zurich, 8093 Zurich, Switzerland*

[3] *Swiss Light Source, Paul Scherrer Institute, CH-5232 Villigen, Switzerland*

[4] *Diamond Light Source, Harwell Campus, Didcot OX11 0DE, United Kingdom*

[5] *Department of Physics, St. Petersburg State University, St. Petersburg, 199034 Russia*

Abstract

We report angle-resolved photoemission experiments resolving the distinct electronic structure of the inequivalent top and bottom (001) surfaces of WTe$_2$. On both surfaces, we identify a surface state that forms a large Fermi-arc emerging out of the bulk electron pocket. Using surface electronic structure calculations, we show that these Fermi arcs are topologically trivial and that their existence is independent of the presence of type-II Weyl points in the bulk band structure. This implies that the observation of surface Fermi arcs alone does not allow the identification of WTe$_2$ as a topological Weyl semimetal. We further use the identification of the two different surfaces to clarify the number of Fermi surface sheets in WTe$_2$.

Main text

Transition metal dichalcogenides (TMDs) have long been studied in many-body physics as model systems for metal insulator transitions, multiband superconductivity and charge density waves [1–3]. In recent years, the interest in TMDs intensified because of the promising optoelectronic properties of monolayer or few-layer devices based on hexagonal semiconducting MX$_2$ compounds with M = W, Mo and X = Se, S [4,5]. Unlike these materials, WTe$_2$ crystallizes in the orthorhombic, non-centrosymmetric 1T' structure (*Pmn2$_1$* space group) and is semi-metallic due to a small overlap of valence and conduction bands at the Fermi level [6,7]. Recent theoretical work [8] predicts that WTe$_2$ is an example of a new type of Weyl semimetal with strongly tilted Weyl cones that arise from topologically protected crossings of valence and conduction bands causing touching points between electron and hole pockets near the Fermi level. In type-I Weyl semimetals, realized for example in TaAs [9–13], the projections of opposite chirality Weyl points onto a surface are isolated from the bulk continuum and must be connected by well-defined Fermi arcs. This is not generally the case for type-II Weyl points, which are necessarily accompanied by bulk carrier pockets. The surface Fermi arcs corresponding to the bulk Weyl points in these materials can emerge within the projection of the bulk carrier pockets, rendering the robust identification of their topological nature challenging. Indeed, very recent angle-resolved photoemission (ARPES) experiments on the related $Mo_xW_{1-x}Te_2$ and $MoTe_2$ systems report conflicting interpretations of the topological character of potential surface states [14–16]. ARPES studies on pure $WTe_2$ have to date not reported any surface states [17–19].

In addition, $WTe_2$ is attracting interest because of its non-saturating magnetoresistance [7] and the recent discovery of pressure induced superconductivity [20,21]. A possible relation between these phenomena and the topological nature of the low-energy surface excitations in $WTe_2$ is an intriguing prospect but has not yet been established. To date, even the basic bulk electronic structure underlying these phenomena remains controversial. The complex magneto-transport properties of $WTe_2$ have successfully been described by a simple two-band model [7,22–25], while quantum oscillation studies have observed at least 7 distinct frequencies that have been rationalized with 4, 6 or 8 closed three-dimensional Fermi surface sheets [26–29]. ARPES experiments, on the other hand, reported 4, 6 or 9 Fermi surface sheets with strongly two-dimensional character [17–19].

In this paper, we report a unifying picture of the surface and bulk electronic structure of $WTe_2$. This is achieved by combining laser and synchrotron based ARPES measurements with surface band structure calculations, allowing us to demonstrate the presence of two distinct surfaces in the *ab*-plane of $WTe_2$ which we associate with the inequivalent top and bottom surfaces of the non-centrosymmetric bulk structure. We identify arc-like surface states with distinct dispersions on the two surfaces and use surface electronic structure calculations to show that their presence is independent of the existence of type-II Weyl points in the bulk. We thus demonstrate that the observation of arc-like surface states does not identify $WTe_2$ as a type-II Weyl semimetal.

Single crystals of $WTe_2$ grown by an optimized chemical vapor transport method [25], were cleaved *in-situ* along the *ab*-plane at temperatures < 20 K, which results in a Te terminated surface. Synchrotron based ARPES experiments were performed at the I05 beamline of Diamond Light Source and the SIS beamline of the Swiss Light Source using photon energies of 40 – 90 eV. Laser-ARPES experiments were performed with a frequency converted diode-laser (LEOS solutions) providing continuous-wave radiation with 206 nm wavelength (hν = 6.01 eV) focused to a spot of ~ 5 μm diameter on the sample surface and an MBS electron spectrometer with a scanning lens system permitting the acquisition of two-dimensional *k*-space maps without rotating the sample.

Measurement temperatures ranged from 6 K to 20 K and the energy and momentum resolution were ~ 15 meV/0.02 Å$^{-1}$ and 2 meV/0.003 Å$^{-1}$ for synchrotron and laser ARPES experiments, respectively. Electronic structure calculations were performed for bulk truncated surfaces with the VASP software package [30] using the PBE functional [31] for exchange-correlation and PAW pseudopotentials [32] that include spin-orbit coupling. Details of the surface state calculations are provided in the Supplemental Material [33].

WTe$_2$ has an orthorhombic unit cell with lattice parameters $a$ = 3.477 Å, $b$ = 6.249 Å and $c$ = 14.018 Å that contains four formula units as illustrated in Fig. 1 [34]. The W atoms form a slightly buckled rectangular lattice with short interatomic distances along the $a$-axis and a larger separation along the $b$-axis and are sandwiched between tellurium layers. The resulting WTe$_2$ sheets are stacked along the $c$ axis with van der Waals interlayer bonding. The only two symmetry operations of the $Pmn2_1$ space group of WTe$_2$ are a mirror symmetry in the $bc$ plane and a glide plane formed by a reflection in the $ac$ plane followed by a translation by (0.5, 0, 0.5). These symmetries result in a non-centrosymmetric structure, which is a prerequisite for the existence of type-II Weyl points in non-magnetic materials [8].

The broken inversion symmetry further implies that the top and bottom Te terminated (001) surfaces indicated by the dashed blue and green lines in Fig. 1a must be inequivalent. However, the differences between the two surfaces are far subtler than in other non-centrosymmetric materials such as BiTeI [35], which might explain why they have been overlooked so far. Clearly, the two inequivalent (001) surfaces have identical chemical composition. As illustrated in Fig. 1(a,b), the outermost WTe$_2$ unit of the two surfaces even shares the buckling patterns of the W and Te layers but differs in the average separation of W and Te planes.

In the following, we discuss ARPES data obtained from the two distinct surfaces. From measurements on ~ 20 different samples we find that the near surface spectral function measured by ARPES can be grouped in two clearly distinct types designated by $A$ and $B$, but shows minimal sample to sample variation within each type. Notably, we observe both surface types in samples from the same batch suggesting that variation in sample composition is not a probable explanation. Moreover, the magnetotransport properties of samples from the same growth batch have been measured by Wang *et. al.* [25] and are similar to those observed by other authors [28] ruling out a sample quality issue. Considering further that WTe$_2$ has a unique cleavage plane in the van der Waals gap, these observations strongly suggest that the two types of electronic structure correspond to the top and bottom (001) surface of WTe$_2$, whose inequivalence is a natural consequence of the broken inversion symmetry. We point out that WTe$_2$ samples have frequent stacking faults and often a substantial mosaic spread, which both limit fully crystalline areas with a single type of (001) surface. This renders it difficult to obtain ARPES data from a single surface termination with conventional experimental setups that average over larger surface areas, which might explain the apparent discrepancy in earlier ARPES studies [17–19]. Indeed, in our ARPES experiments with a highly focused laser beam probing a small surface area only, we find that the prevalence of each type of surface is close to 50%, while the cleaved surface of larger samples often shows both surface types.

Representative Fermi surfaces of A and B type cleaved surfaces obtained by laser-ARPES are shown in the second column of Fig. 1. On both surfaces we find closed electron and hole pockets in agreement with earlier reports [17–19]. Their contours are summarized in Fig. 1 (e,f) by blue and green areas. The most striking difference between the two surface types is the dispersion of an arc-like surface state emerging out of the electron pocket shown in red in Fig. 1 (e,f). On the B-type surface, the Fermi arc appears to connect electron and hole pockets while on the A-type surface, it barely splits off the closed electron pocket. These observations are in qualitative agreement with the surface density of states calculated for the two surface terminations of $WTe_2$ shown in Fig. 1(g,h). However, on a quantitative level there are clear differences between experiments and calculations. Notably, the bulk-like electron and hole pockets are much larger in the calculation while the difference in surface state dispersion between A and B type surface is more pronounced in the experiment. The latter might arise from surface relaxation, which is not taken into account in the calculations. As we will discuss later, these quantitative differences prohibit a reliable determination of the topological properties of $WTe_2$. On the other hand, they do not affect the identification of surface states and bulk-like states from a comparison of the ARPES data with the band structure calculations.

In Fig. 2(a,b) we show band dispersion plots along ΓX obtained on type-A and type-B surfaces respectively with synchrotron radiation at photon energies that probe the electronic structure near the $k_z$ = 0 plane where band structure calculations predict the maximal extension of bulk electron and hole pockets. Laser-ARPES measurements along the same direction are presented in Figs. 2(e,f). Although photon energy dependent matrix elements modulate the spectral weight, we find that the band dispersion at the photon-energies shown here is similar. We thus comprehensively characterize the surface and bulk electronic structure using peaks extracted from both laser-ARPES and synchrotron measurements with different polarizations. The resulting band dispersion along ΓX is shown by black symbols in Figs. 2(e) and 2(f) together with colored guides-to-the-eye indicating surface states in red, a bulk electron pocket in blue and two hole pockets in green. The corresponding bulk band structure calculation along the ΓX direction is displayed in Fig. 2(g) using the same color code. From the good qualitative agreement with the data we conclude that no bulk bands evade detection in our experiment. The calculations further show a small lifting of the two-fold degeneracy of the electron and hole pockets which is only resolved experimentally in limited $k$-space areas and arises from the inequivalent W sites in the 1T' structure. Hence, our ARPES data correspond to a total number of four inequivalent hole pockets and two electron pockets in half of the Brillouin zone corresponding to positive values of $k_x$, with the same number of symmetry-equivalent pockets present at negative $k_x$ values. From the shaded areas in Fig.1(e) we estimate cross-sections in the $ab$ plane of 0.015(3) Å$^{-2}$ for the nearly degenerate electron pockets, and of 0.018(4) Å$^{-2}$ and 0.011(2) Å$^{-2}$, for the split sheets of the large hole pocket. These Fermi surface areas are in good agreement with the quantum oscillation frequencies observed in Ref. [26]. Additionally, we observe two nearly degenerate small hole pockets (light-green) with areas of 0.004(1) Å$^{-2}$ that have not been observed in quantum oscillation experiments indicating that the near surface electronic structure probed by ARPES differs slightly from the bulk. We note that all these Fermi surface sheets are fully spin-polarized, which is a natural consequence of the broken inversion symmetry of WTe$_2$ and the large spin-orbit interaction of W. This restricts the phase space for back-scattering and was invoked to explain the unusually large magnetoresistance of WTe$_2$ [18]. Conversely, it prohibits an identification of the surface state topology from spin-resolved experiments.

In Fig. 2 (h-j) we clarify the dimensionality of the electronic states and thus their surface or bulk character. Given the layered structure of WTe$_2$ one might naively expect that the entire electronic structure should be largely two-dimensional. However, magnetotransport experiments have unambiguously shown that the bulk electron and hole pockets form closed Fermi surface sheets with finite extension along $k_z$ [26] in apparent contrast with published ARPES data [18]. To resolve this discrepancy, we measured the out-of-plane dispersion on both surface types by varying the excitation energy between 40 eV and 90 eV. In Fig. 2(h,j) we display the resulting $e(k_z)$ dispersion plots on the B-type surface for two different $k_x$ values indicated by dashed lines in panel (f). At $k_x$ corresponding to the bottom of the electron pocket, we find a clear dispersion of the electronic states demonstrating that they arise from a closed 3D electron pocket. From fits that include a Lorentzian $k_z$-broadening reflecting the small probing depth of ARPES, we estimate the Fermi wave vectors for the electron pockets along z to be in the range of $k_F$ = 0.07 – 0.13 Å$^{-1}$, in fair agreement with the values deduced from quantum oscillation experiments [26]. Conversely, Fig. 2(j) shows that the surface state identified from the comparison with the band structure calculations indeed does not disperse with $k_z$ over an entire Brillouin zone and is thus fully 2D. Fig. 2(i) shows the Fermi surface in the $k_x – k_z$ plane for the type-A surface. One readily identifies two straight lines at the edge of the bulk electron and hole pockets, which must arise from 2D states that do not disperse in the third dimension. We identify these states with the surface states SS1 and SS2 found in our band structure calculations shown in Fig. 3. Hence, these measurements confirm the identification of the surface state SS1 on the A-type surface in immediate vicinity of the bulk electron pocket. This surface state is thus present on both surface types and can be distinguished from the three dimensional bulk bands of $WTe_2$.

For the remainder of this letter, we discuss the relation of the surface state Fermi arcs identified in our work with the topological properties of $WTe_2$. Using the experimental low-temperature crystal structure, Ref. [8] predicted the existence of eight type-II Weyl points in $WTe_2$ and the appearance of topological surface states producing topological Fermi arcs on the (001) surface connecting the projections of neighboring Weyl points of opposite chirality. Ref. [8] also predicted that a $k_y$ = 0 cut of the surface Brillouin zone can only cross an even number of Kramer's pairs of surface states. This is indeed the case, as shown in Fig. 3(a): Two surface states (SS1, SS2) traversing the gap between valence and conduction bands are visible in the figure (another two are located at -$k_x$). The presence of two surface states can be attributed to the two sets of Weyl points present in the system. Analogously to the case of $MoTe_2$ where only 4 Weyl points are predicted to exist [36], each set of 4 Weyl points creates a surface state that is topological. However, in the absence of additional topological protection, the two states that appear from the two sets can hybridize and form a state that is trivial over most of *k*-space except for the short section between adjacent Weyl points. Given the qualitative agreement of our ARPES data with the numerical calculation of the (001) surface density of states finding 8 type-II Weyl points (Fig. 1(g,h)) it is tempting to propose such a scenario, where the arc seen in experiment is trivial and explained as arising from one of the two surface states created by a set of 4 type-II Weyl points.

On the other hand, it is known that the electronic state identified in Ref. [8] is delicate with Weyl points of opposite chirality located in immediate vicinity of each other in the Brillouin zone. They are therefore easily annihilated by small relative shifts in the dispersion of the corresponding bulk bands. Moreover, as shown in Refs. [8,36], small changes to the lattice structure, even those induced by temperature variations, can change the particular arrangement and the total number of Weyl points in band structure calculations of $WTe_2$ and related compounds. It is thus important to test the robustness of the interpretation of the experimental findings. To this end, we show in Fig. 3 surface band structure calculations for $WTe_2$ calculated for the slightly different lattice constants found at room temperature and low-temperature [6,34]. Clearly, the surface density of states including the dispersion of the surface states SS1 and SS2 is virtually identical for the two calculations and the minute differences are well below the resolution of the best contemporary ARPES experiments. Remarkably though, a careful inspection of the bulk band structure shows that all 8 type-II Weyl points found for the low-temperature lattice parameters are annihilated in the calculation for the slightly larger room temperature unit cell. The latter calculation therefore corresponds to a topologically trivial semimetal state. This clearly shows that the observation of the large Fermi arc SS1 does not identify $WTe_2$ as a type-II topological Weyl semimetal since the presence of the associated Fermi arc does not imply the existence of type-II Weyl points. We stress that any robust identification of a type-II Weyl semimetal state needs to exclude all plausible topologically trivial scenarios. Our identification of a topologically trivial scenario presented in Fig. 3(b), which provides an equally good description of the data as the topologically non-trivial calculation of Fig 3(a), renders this task nearly impossible in the case of $WTe_2$.

In conclusion, we used small-spot laser-ARPES with a scanning electron lens permitting the acquisition of very high-resolution Fermi surface maps from surface areas with ≈ 5 μm lateral dimension to resolve the distinct electronic structure of the two inequivalent (001) surfaces of $WTe_2$. This allowed us to identify surface state Fermi arcs emerging out of the bulk electron pocket. Within our surface electronic structure calculations, these Fermi arcs are found to be topologically trivial in all scenarios. Moreover, they even persist for topologically trivial bulk electronic structures. We expect that the new experimental capabilities used here will prove useful in the search for more robust type-II Weyl semimetals in other non-centrosymmetric materials, whose top and bottom surfaces are necessarily inequivalent.

After submission of our work, three other groups confirmed the presence of the large surface state Fermi arc reported here and interpreted this observation as evidence for a type-II Weyl semimetal state [37–39]. Fig. 3 of our paper shows that this conclusion is premature.

We gratefully acknowledge discussions with B. A. Bernevig, A. Morpurgo, I. G. Lezama and L. Wang. This work was supported by the Swiss National Science Foundation (200021-146995) and the Ambizione grant (PZ00P2_161327). Q.S.W. and A.A.S. acknowledge funding from Microsoft Research and the Swiss National Science Foundation through the National Competence Centers in Research MARVEL and QSIT. The authors acknowledge Diamond Light Source for time on beamline I05 under proposal SI12404.

Figure Captions

FIG. 1. (a) Side view of the crystal structure of $WTe_2$. (b) Top and bottom view of the $WTe_2$ lattice. The + and - (U and D) symbol denotes the Te (W) atoms that are further away and closer to the mean position. (c) and (d) ARPES Fermi surface of $WTe_2$ obtained with 6.01 eV excitation energy measured on two distinct surfaces parallel to the ab-plane denoted as type A and type B, respectively. (e) and (f) Fermi surface contours extracted from measurements on samples type A and B respectively. Measurements obtained with different polarizations are shown in different symbols, shaded blue and green areas denote the electron and hole pockets and the surface state SS1 is shown in red. (g) and (h) Fermi surface calculated for a semi-infinite crystal for the inequivalent top and bottom surfaces.

FIG. 2. (a-c) Dispersion of the low-energy electronic states measured along the Γ-X direction on a type A surface. (b-d) Same measurements on type B surface. (e) and (f) extracted dispersion of the bands

from measurements with different excitation energies and polarizations for surfaces of type A and B respectively. The thick lines are guide to the eyes. (g) Bulk band structure of $WTe_2$ along Γ-X from density functional theory. (h-j) kz dispersion of the bulk electron pocket and surface state on a type B surface measured at kx = 0.36 Å-1 and kx = 0.3 Å-1 respectively. (i) Constant energy surface measured at $E_F$ on a type A surface.

FIG. 3. (a,b) momentum resolved (001) surface density of states calculated along the Γ-X direction for the bottom surface of $WTe_2$. The lattice parameters at T=113 K and room temperature are used in (a) and (b) respectively resulting in the presence of 8 and 0 Weyl points. (c) Fermi surface calculation of the $WTe_2$ bottom surface. The top and bottom half correspond to the case of 8 and 0 Weyl points respectively.

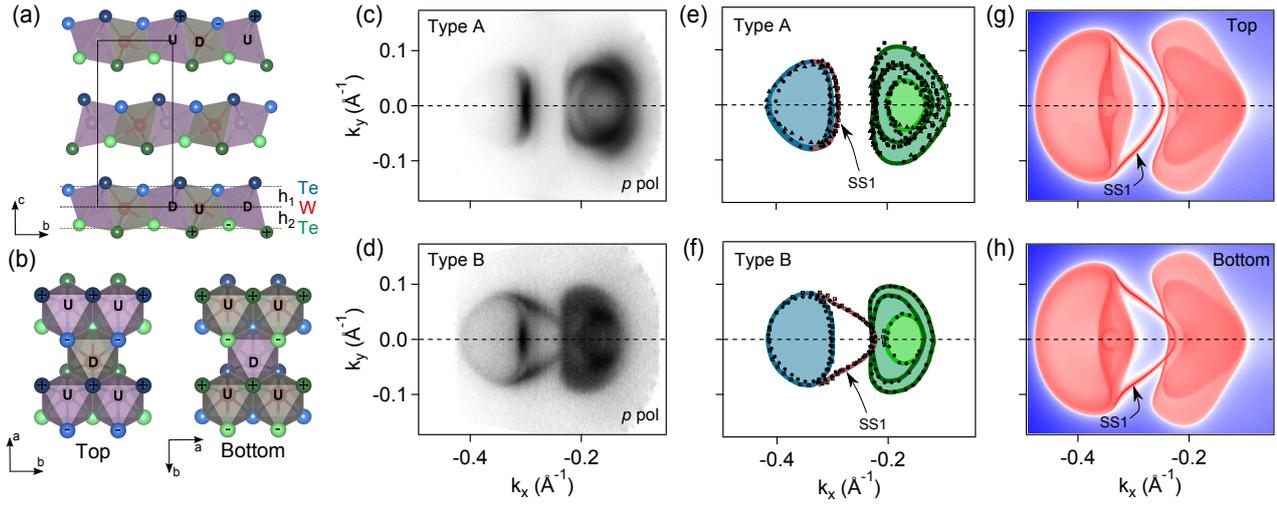

FIG. 1. (a) Side view of the crystal structure of WTe$_2$. (b) Top and bottom view of the WTe$_2$ lattice. The + and - (U and D) symbol denotes the Te (W) atoms that are further away and closer to the mean position. (c) and (d) ARPES Fermi surface of WTe$_2$ obtained with 6.01 eV excitation energy measured on two distinct surfaces parallel to the ab-plane denoted as type A and type B, respectively. (e) and (f) Fermi surface contours extracted from measurements on samples type A and B respectively. Measurements obtained with different polarizations are shown in different symbols, shaded blue and green areas denote the electron and hole pockets and the surface state SS1 is shown in red. (g) and (h) Fermi surface calculated for a semi-infinite crystal for the inequivalent top and bottom surfaces.

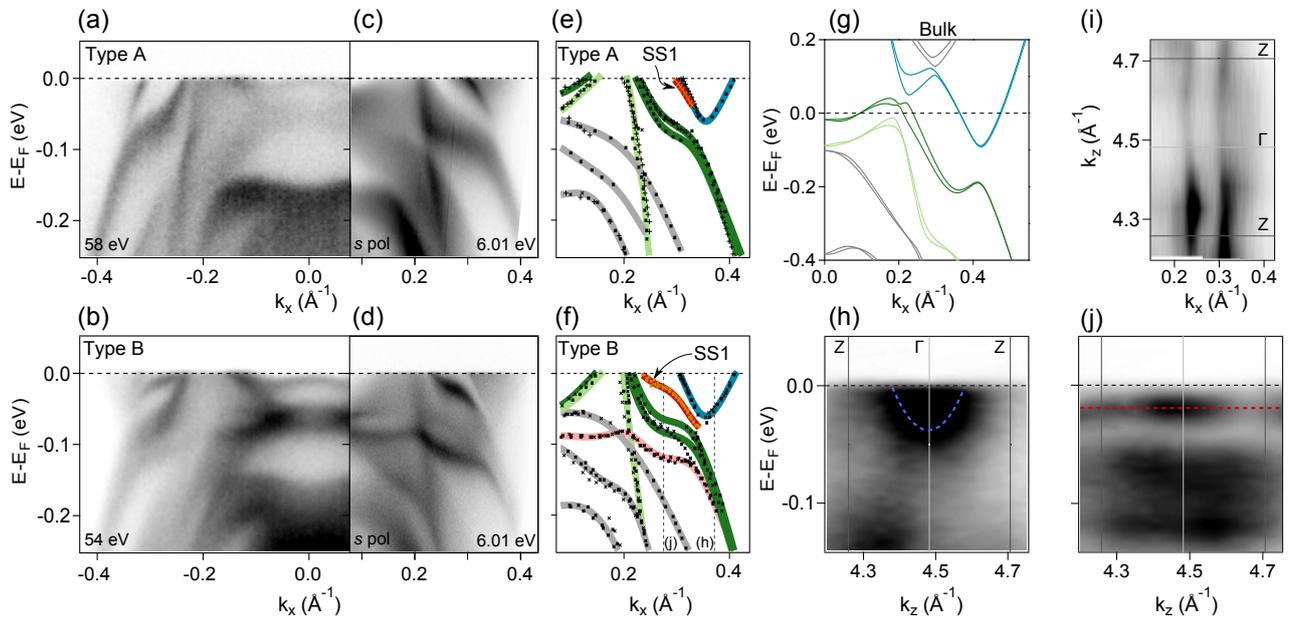

FIG. 2. (a-c) Dispersion of the low-energy electronic states measured along the Γ-X direction on a type A surface. (b-d) Same measurements on type B surface. (e) and (f) extracted dispersion of the bands from measurements with different excitation energies and polarizations for surfaces of type A and B respectively. The thick lines are guide to the eyes. (g) Bulk band structure of $WTe_2$ along Γ-X from density functional theory. (h-j) $k_z$ dispersion of the bulk electron pocket and surface state on a type B surface measured at $k_x = 0.36$ Å$^{-1}$ and $k_x = 0.3$ Å$^{-1}$ respectively. (i) Constant energy surface measured at $E_F$ on a type A surface.

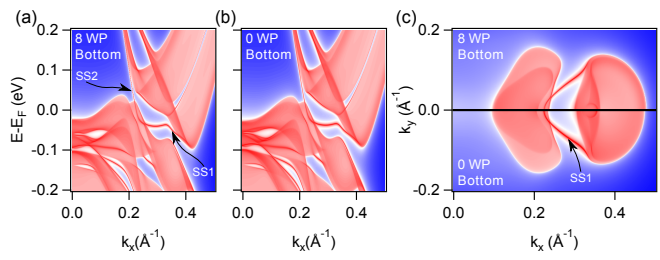

FIG. 3. (a,b) momentum resolved (001) surface density of states calculated along the Γ-X direction for the bottom surface of WTe$_2$. The lattice parameters at T=113 K and room temperature are used in (a) and (b) respectively resulting in the presence of 8 and 0 Weyl points. (c) Fermi surface calculation of the WTe$_2$ bottom surface. The top and bottom half correspond to the case of 8 and 0 Weyl points respectively.